\documentclass[letterpaper, 10pt, conference]{ieeeconf}
\IEEEoverridecommandlockouts
\usepackage{amssymb}
\usepackage{amsmath}
\usepackage{cite}
\usepackage[utf8]{inputenc}
\usepackage{graphicx}
\usepackage{hyperref}
\usepackage{xcolor}
\usepackage{algorithm}
\usepackage{algpseudocode}

\title{\LARGE \bf Analyzing Fitts' Law using Offline and Online Optimal Control with Motor Noise}
\author{Riley Bridges, Ethan Parham, Jing Shuang (Lisa) Li 
\thanks{R.B. and E.P. were with the Department of Electrical Engineering and Computer Science and Department of Mechanical Engineering (respectively) at the University of Michigan when the bulk of this work was conducted. J.S.L. is with the Department of Electrical Engineering and Computer Science at the University of Michigan. {\tt\small rlybrdgs@umich.edu, ejparham@umich.edu, jslisali@umich.edu}}
}

\begin{document}

\maketitle
\thispagestyle{plain}
\pagestyle{plain}

\begin{abstract}
The cause of the speed-accuracy tradeoff (typically quantified via Fitts' Law) is a debated topic of interest in motor neuroscience, and is commonly studied using tools from control theory. Two prominent theories involve the presence of signal dependent motor noise and planning variability --- these factors are generally incorporated separately. In this work, we study how well the simultaneous presence of both factors explains the speed-accuracy tradeoff. A human arm reaching model is developed with bio-realistic signal dependent motor noise, and a Gaussian noise model is used to deterministically approximate the motor noise. Both offline trajectory optimization and online model predictive control are used to simulate the planning and execution of several different reaching tasks with varying target sizes and movement durations. These reaching trajectories are then compared to experimental human reaching data, revealing that both models produce behavior consistent with humans, and the speed-accuracy tradeoff is present in both online and offline control. These results suggest the speed-accuracy tradeoff is likely caused by a combination of these two factors, and also that it plays a role in both offline and online computation.
\end{abstract}

\section{Introduction}

The speed-accuracy tradeoff is a phenomenon in motor neuroscience which describes the inverse relationship between the speed of a motion and the required accuracy of that motion.
The presence of the speed-accuracy tradeoff means that the more accuracy a movement requires, the more time the movement will take and vice versa.
It is fairly ubiquitous across different animal species as well as different task types \cite{zimmerman_book}, however it is frequently studied specifically as a property of human motor tasks. Understanding the speed-accuracy tradeoff and its causes is an important part of understanding human motion, as it is a fundamental aspect of human motor control. For this reason, it is often studied using tools from control theory.
The speed-accuracy tradeoff can be quantified using Fitts' law, which relates the width of a target of motion $W$, the distance of the movement $A$, and the duration of the movement $MD$. 


\begin{equation}
    MD = a + b \log_2 \left(\frac{2A}{W}\right) \label{eq:fitts_law}
\end{equation}

The causes of the speed-accuracy tradeoff are not well agreed upon. Historically, the speed-accuracy tradeoff has been attributed to signal dependent motor noise, which is noise that increases in variance as the strength of the neural control signal increases \cite{signal_dependent_motor_noise}. 
However, recent work has shown that the speed-accuracy tradeoff can be reproduced without the presence of motor noise. Trajectory optimization was used to model the planning of a reaching motion, and the speed-accuracy tradeoff was reproduced as the result of motor planning variability \cite{original_paper_high_fidelity}. This is where the brain is theorized to be optimizing the trajectory of a motion, and higher accuracy requirements make the optimization more likely to fall into a suboptimal local minimum.

Several previous works have studied optimal control of a human arm in the presence of motor noise. In \cite{ilqg}, an iterative linear quadratic gaussian controller was studied for executing reaching motions with motor noise.
This approach was further iterated on in \cite{stochastic_model}, where a similar human arm model was controlled using a combination of feedforward trajectory optimization and full state optimal feedback control. These works did not, however, study the speed-accuracy tradeoff, and the motor noise in the model was not signal dependent. In \cite{cost_benefit_tradeoff}, an optimal controller was used to study the speed-accuracy tradeoff in the presence of signal dependent motor noise. However, the cost benefit tradeoff was the primary focus of the work, it did not compare feed forward and feedback control, and it used a learning based stochastic optimization method rather than a deterministic optimization method.

We aim to expand on this line of study by incorporating both signal dependent motor noise and motor planning variability into a human reaching model, and by comparing the effects of offline and online optimal control using a deterministic constrained optimization method.

The remainder of the paper is organized as follows. In section II we introduce the system model and problem formulation. In section III we derive the formulations for both feed forward planning and feedback control. In section IV we present the results of our simulations and discuss their implications. Finally, in section V we discuss limitations and future work.

\section{Problem Formulation}

We aim to answer the following questions: 
\begin{itemize}
    \item When using trajectory optimization to plan reaching movements on a human arm model, how well does the modeled behavior match human behavior, particularly with respect to the speed-accuracy tradeoff?

    \item How does the introduction of signal-dependent motor noise to the model affect this comparison?

    \item How does this behavior change when simulating execution of reaching movements using feedback control rather than just planning them?
\end{itemize}
For each of these questions, we use the following metrics to compare the simulated behavior (collected from our models) to human behavior (experimental data collected and used in \cite{original_paper_high_fidelity}):
\begin{itemize}
    \item Similarity of hand velocity profile during center-out fast reaching movements, evaluated qualitatively by appearance and quantitatively by time and value of maximum velocity:
    \begin{align}
        V_{\text{max}} &= \max \left|V_{\text{sim}}(t)\right| \\
        T_{\text{max}} &= \arg\max_{t \in [0, t_f]} \left|V_{\text{sim}}(t)\right|
    \end{align}
    where $V_{\text{sim}}$ is the simulated velocity of the hand during its reaching trajectory, and $t_f$ is the time at which the hand reaches the target.

    \item Delay in time of maximum velocity between large and small targets:
    \begin{multline}
        t_{\text{delay}} = \arg\max_{t \in [0, t_f]} \left|V_{\text{sim}}^{L}(t)\right| - \arg\max_{t \in [0, t_f]} \left|V_{\text{sim}}^{S}(t)\right|
    \end{multline}
    where $V_{\text{sim}}^{L}$ and $V_{\text{sim}}^{S}$ are the simulated velocities of the hand during its reaching trajectory to large and small targets, respectively.
    \item Fitts' Law model parameters $a$ and $b$.
    Several reaching tasks with variable target distance and target width will be simulated. A least squares regression will then be performed on the resulting movement duration data from each trial in order to fit the parameters:
    \begin{align}
        \begin{bmatrix}
            a^* \\ b^*
        \end{bmatrix} &= \arg\min_{a, b} \sum_{i=1}^N \left(MD_i - \left(a + b \log_2\left(\frac{2A_i}{W_i}\right)\right)\right)^2
    \end{align}
    where $MD_i$ is the movement duration, $A_i$ the target distance, and $W_i$ the target width for the $i$th trial, and $N$ is the number of trials.
\end{itemize}

\subsection{System Model and Dynamics}
\begin{figure}[h]
    \centering
    \includegraphics[width=0.5\textwidth]{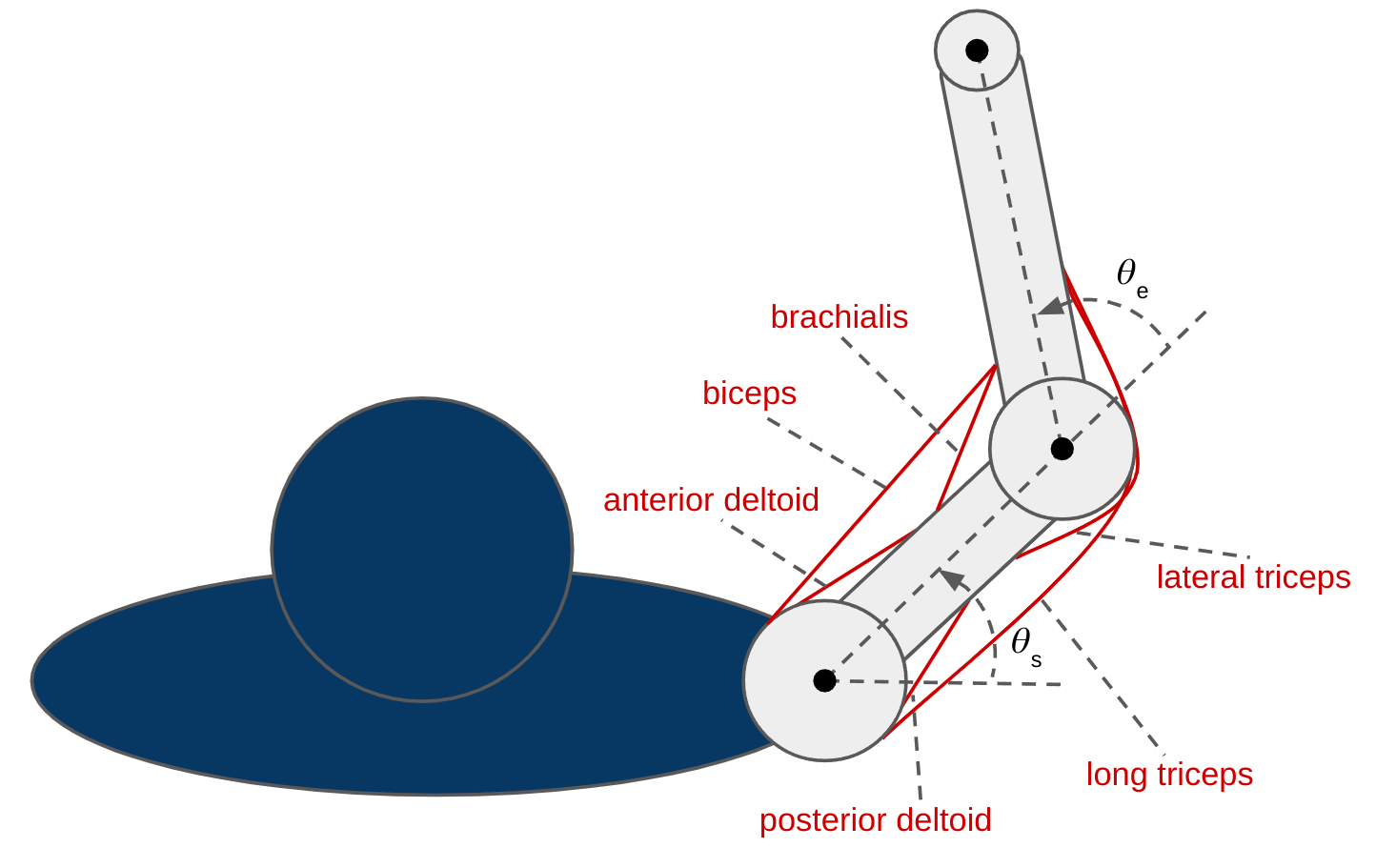}
    \caption{Diagram of the musculoskeletal human arm model. }
    \label{fig:arm_model}
\end{figure}

We model the human arm as a two degree of freedom planar arm with shoulder and elbow joints, actuated by six Hill-type muscles. The majority of this model is taken from \cite{stochastic_model}, which we will now review here. Adaptations have been made from the original model to include signal dependent motor noise and to formulate it as a constrained optimization problem. A diagram of the model is shown in figure \ref{fig:arm_model}. 

The state and control input vectors are defined below:
\begin{align}
    \mathbf{x} &= \begin{bmatrix}
        \mathbf{q} &
        \dot{\mathbf{q}}
    \end{bmatrix}^\top = \begin{bmatrix}
        \theta_s &
        \theta_e &
        \dot{\theta}_s &
        \dot{\theta}_e
    \end{bmatrix}^\top \\
    \mathbf{u} &= \begin{bmatrix}
        a_{\text{brach}} & a_{\text{lattri}} & a_{\text{antdel}} & a_{\text{postdel}} & a_{\text{bic}} & a_{\text{lattri}}
    \end{bmatrix}
\end{align}
where $\theta_s$ and $\theta_e$ are the shoulder and elbow joint angles, respectively, and $\dot{\theta}_s$ and $\dot{\theta}_e$ are the corresponding joint angular velocities. The control input $\mathbf{u}$ is a vector of muscle activations for each of the 6 muscles in the model. The stochastic dynamics of the system are given by the following differential equation:
\begin{align}
    \dot{\mathbf{x}} &= f(\mathbf{x}, \mathbf{u}, \mathbf{w}) \quad \mathbf{w} \sim \mathcal{N}(0, \Sigma_w) \\
    &= \begin{bmatrix}
        \dot{\mathbf{q}} \\
        M(\mathbf{q})^{-1} \left(C(\mathbf{q}, \dot{\mathbf{q}}) + T_M(\tilde{\mathbf{u}}, \mathbf{q})\right)
    \end{bmatrix} \\
    \tilde{\mathbf{u}} &= \mathbf{u} + \text{diag}(\mathbf{u}) \cdot \mathbf{w}
\end{align}
where $M(\mathbf{q})$ is the mass matrix of the arm model, $C(\mathbf{q}, \dot{\mathbf{q}})$ is the term describing the Coriolis forces, and $T_M(\mathbf{u}, \mathbf{q})$ is the term describing the torques generated by the muscles.
Note that the noise vector is effectively scaled element-wise by the control input vector, thus modeling signal dependent motor noise as described earlier.

We can then approximate the stochastic state trajectories of the system as normally distributed trajectories, allowing us to represent the stochastic trajectory as a mean trajectory $\mathbf{\bar{x}}(t) \in \mathbb{R}^4$ and a covariance trajectory $P(t) \in \mathbb{R}^{4\times 4}$. We can then use this representation to create a deterministic first order approximation of the dynamics: 
\begin{align*}
    \mathbf{\dot{\bar{x}}}(t) &= f(\mathbf{\bar{x}}(t), \mathbf{u}(t), 0) \\
    \dot{P}(t) &= A(t)P(t) + P(t)A(t)^\top + C(t) \Sigma_w C(t)^\top \\
    &\quad A(t) = \frac{\partial f}{\partial \mathbf{x}}\bigg|_{\mathbf{\bar{x}}(t), \mathbf{u}(t), 0} \\
    &\quad C(t) = \frac{\partial f}{\partial \mathbf{w}}\bigg|_{\mathbf{\bar{x}}(t), \mathbf{u}(t), 0}
\end{align*}

\subsection{Trajectory Optimization}
To plan the optimal arm trajectory for a given reaching task, we aim to solve the following optimal control problem:
\begin{align}
    \mathbf{u}^* = \arg\min_{\mathbf{u}(t)} &\quad k_u \cdot \int_0^{t_f} \mathbf{u}(t)^\top \mathbf{u}(t) dt + k_t \cdot t_f \label{eq:cost} \\
    \text{subject to} &\quad \mathbf{\dot{\bar{x}}}(t) = f(\mathbf{\bar{x}}(t), \mathbf{u}(t), 0), \quad \mathbf{\bar{x}}(0) = \mathbf{x}_0 \label{eq:dynamics_constraint} \\
    \begin{split}
    &\quad \dot{P}(t) = A(t)P(t) + P(t)A(t)^\top \\
    &\quad \quad + C(t) \Sigma_w C(t)^\top, \quad P(0) = P_0 
    \end{split}\label{eq:dynamics_constraint_p} \\
    &\quad \mathbf{u}(t) \in (0.001, 1), \quad \mathbf{x}(t) \in (0, \pi) \label{eq:constraints} \\
    &\quad \mathbf{\dot{\bar{x}}}(0) = 0, \quad \mathbf{\dot{\bar{x}}}(t_f) = 0, \quad t_f > 0 \label{eq:boundary_constraints} \\
    &\quad F_k(\mathbf{\bar{x}}(t_f)) = \begin{bmatrix} \mathbf{p}_{\text{target}} & 0 \end{bmatrix}^\top \label{eq:target_constraint} \\
    \begin{split}
    &\quad [HP(t_f)H^\top]_{i,i} \leq \sigma_{\text{target}, i}^2 \ , \\
    &\quad H = \frac{\partial F_k}{\partial \mathbf{x}}\bigg|_{\mathbf{\bar{x}}(t_f)}, \quad i = 1, \ldots, n 
    \end{split}\label{eq:target_variance_constraint}
\end{align}
where $t_f$ is the duration of the trajectory, $k_u$ is a weight for muscle activation, $k_t$ is a weight for duration, $\mathbf{x}_0$ is the initial state of the arm, $P_0$ is the initial state covariance, $F_k: \mathbb{R}^4 \to \mathbb{R}^4$ is the arm forward kinematics function, $\mathbf{p}_{\text{target}}$ is the position of the target, and $\sigma_{\text{target}, i}$ is the desired standard deviation of the final position of the $i$th dimension of the target.

Equation \ref{eq:cost} works to minimize muscle activations and trajectory duration. Equations \ref{eq:dynamics_constraint} and \ref{eq:dynamics_constraint_p} enforce the dynamics of the system. Equation \ref{eq:constraints} enforces muscle activation and joint bounds. Equation \ref{eq:boundary_constraints} enforces zero velocity and acceleration at the beginning and end of the trajectory, while equation \ref{eq:target_constraint} enforces the final mean position of the arm's end effector to be at the target. Finally, equation \ref{eq:target_variance_constraint} enforces the final position of the arm's end effector to be within a certain standard deviation of the target position. This standard deviation can be adjusted to model variation in target size in order to test the speed-accuracy tradeoff.




\section{Approach}

\subsection{Offline Planning using Direct Collocation}
To solved the trajectory optimization problem, we adapted the optimal control framework from \cite{stochastic_model} to perform only feed forward control and to fit our adapted problem formulation. This implementation required us to discretize the muscle activation trajectory into a vector of $N$ nodes, and to optimize over that vector. To enforce the dynamics efficiently, we use a direct collocation based method \cite{direct_collocation}. This involves similarly discretizing the state trajectory into a series of $N$ nodes that are added to the design vector, and then enforcing the dynamics constraints between each consecutive node. This formulation results in the following design vector:
\begin{align}
    \mathcal{X} &= \begin{bmatrix}
        \mathbf{u}_1  \cdots \mathbf{u}_N & \mathbf{\bar{x}}_1 \cdots \mathbf{\bar{x}}_N & P_1 \cdots P_N & t_f
    \end{bmatrix}^\top
\end{align}
where $\mathbf{u}_i$, $\mathbf{\bar{x}}_i$, and $P_i$ are the muscle activations, state, and state covariance at the $i$th node, respectively. Then the discretized optimization problem can be formulated as follows:
\begin{align}
    \mathcal{X}^* = \arg\min_{\mathcal{X}} &\quad k_u \cdot \sum_{i=1}^{N} \mathbf{u}_i^\top R \mathbf{u}_i + k_t \cdot t_f \label{eq:dcost} \\
    \text{subject to} &\quad \mathbf{\bar{x}}_{i+1} = \mathbf{\bar{x}}_i + f(\mathbf{\bar{x}}_i, \mathbf{u}_i, 0) \delta t \\
    \begin{split}
        &\quad P_{i+1} = (I + A_i \delta t)P_i(I + A_i \delta t)^\top \\
        &\quad \quad + C_i \Sigma_w C_i^\top \delta t
    \end{split} \label{test} \\
    &\quad \mathbf{u}_i \in (0.001, 1), \quad \mathbf{x}_i \in (0, \pi) \\
    &\quad \mathbf{\bar{x}}_1 = \mathbf{x}_0, \quad f(\mathbf{\bar{x}}_0, \mathbf{u}_0, 0) = 0, \quad t_f > 0 \\
    &\quad f(\mathbf{\bar{x}}_N, \mathbf{u}_N, 0) = 0, \quad F_k(\mathbf{\bar{x}}_N) = \mathbf{p}_{\text{target}} \label{eq:dtarget_constraint} \\
    &\quad [HP_NH^\top]_{i,i} \leq \sigma_{\text{target}, i}^2, \quad H = \frac{\partial F_k}{\partial \mathbf{x}}\bigg|_{\mathbf{\bar{x}}_N} \\
    &\quad i = 1, \ldots, n \label{eq:dtarget_variance_constraint}
\end{align}

Note that as in \cite{stochastic_model}, the dynamics constraints are further modified using a positive definiteness preserving discretization to minimize numerical errors.

\begin{figure}
    \centering
    \includegraphics[width=0.45\textwidth]{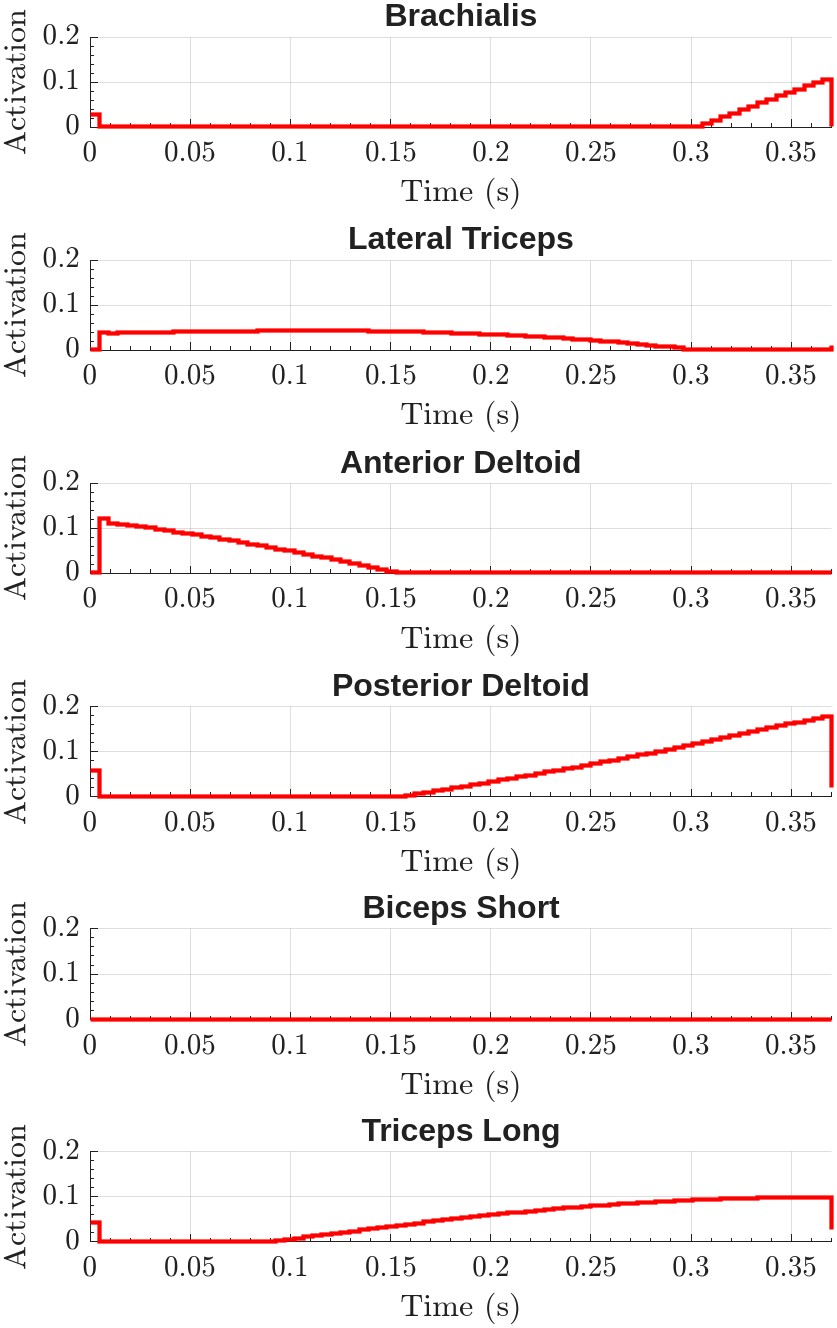}
    \caption{Muscle activation trajectories for an example reaching task.}
    \label{fig:demo_activations}
\end{figure}

\begin{figure}
    \centering
    \includegraphics[width=0.45\textwidth]{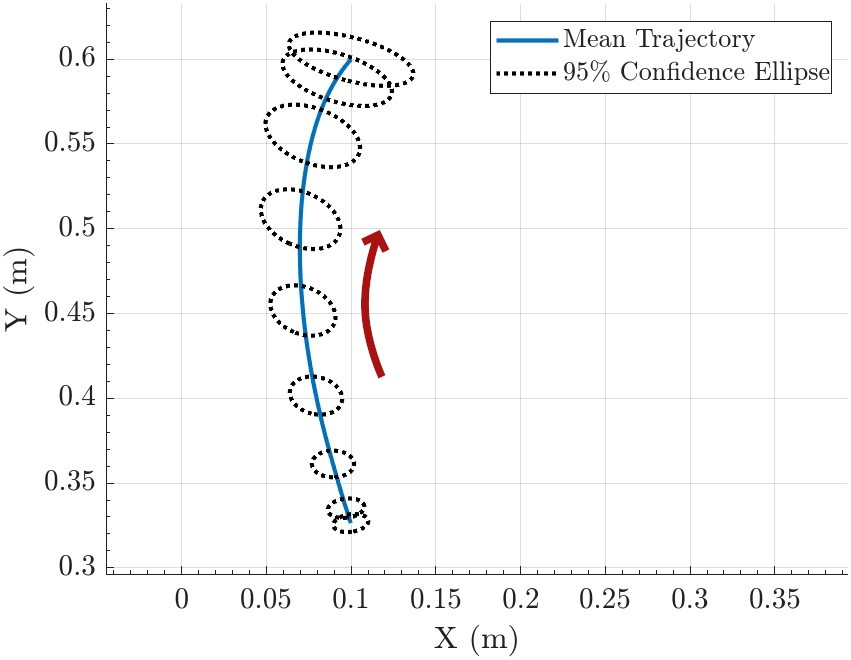}
    \caption{End Effector trajectory for wn example reaching task.}
    \label{fig:demo_position}
\end{figure}

Fig. \ref{fig:demo_activations} shows example muscle activation trajectories associated with a specific reach, and Fig. \ref{fig:demo_position} shows the corresponding end-effector trajectory.

\subsection{Model Predictive Feedback Control}
To simulate the act of executing a reaching motion, rather than just planning it, we implemented a nonlinear model predictive controller (MPC). This controller uses the same nonlinear trajectory optimization problem as the offline planner, and simply solves that problem at each iteration of the controller. The control algorithm is detailed in Algorithm \ref{alg:mpc}.

\begin{algorithm}
\caption{Model Predictive Feedback Simulation}\label{alg:mpc}
 \hspace*{\algorithmicindent} \textbf{Input} \\
    \hspace*{\algorithmicindent} \quad $\mathbf{x}_0$: initial arm state \\
    \hspace*{\algorithmicindent} \quad $t_{\text{iter}}$: duration to execute before replanning \\
    \hspace*{\algorithmicindent} \quad $\mathbf{p}_{\text{target}}$: target position \\
    \hspace*{\algorithmicindent} \quad $\sigma_{\text{target}}$: target standard deviation \\
    \hspace*{\algorithmicindent} \quad $N$: number of nodes in trajectory 
\begin{algorithmic}
\State $\mathbf{x}_{\text{current}} \gets \mathbf{x}_0$
\While{$\|F_k(\mathbf{x}_{\text{current}}) - \mathbf{p}_{\text{target}}\| > \sigma_{\text{target}}$}
    \State $\mathbf{u}^*, \delta t \gets \text{SolveOptimization}(\mathbf{x}_{\text{current}}, \mathbf{p}_{\text{target}}, \sigma_{\text{target}}, N)$
    \State $N_{\text{iter}} \gets \min(\lceil t_{\text{iter}} / \delta t \rceil, N)$
    \For{$i = 1$ to $N_{\text{iter}}$}
        \State $\mathbf{w} \gets \mathcal{N}(0, \Sigma_w)$
        \State $\mathbf{x}_{\text{current}} \gets \mathbf{x}_{\text{current}} + f(\mathbf{x}_{\text{current}}, \mathbf{u}_i^*, \mathbf{w}) \delta t$
    \EndFor
\EndWhile
\end{algorithmic}
\end{algorithm}

\subsection{Implementation}
To execute both our offline planning and feedback MPC simulations, the algorithms were implemented in MATLAB. 
We used the CasADi library \cite{casadi} as an optimization framework to specify the problem and provide automatic differentiation for gradient and Jacobian computation. The IPOPT nonlinear solver was then used to perform the numerical optimization, with trivial constants provided as initial guesses for elements of the design vector. 
IPOPT was run for a maximum of 3000 iterations, after which the problem was considered infeasible. For MPC simulation, the optimization at the first iteration was solved as described previously, and all consecutive iterations were ``warm-started'' by setting the initial design vector guess equal to the solution from the previous iteration. Most optimization runs converged in less than 500 iterations and took less than 30 seconds to run on a high end consumer grade CPU, although MPC optimization runs that were "warm started" typically converged significantly faster and in fewer iterations. All of the code used to implement the optimization problem can be found in the project GitHub repository: \href{https://github.com/rbridges12/speed-accuracy-optimal-control-paper}{https://github.com/rbridges12/speed-accuracy-optimal-control-paper}.

\section{Simulations}
\subsection{Model Validation}

\begin{figure}[h]
    \centering
    \includegraphics[width=1\linewidth]{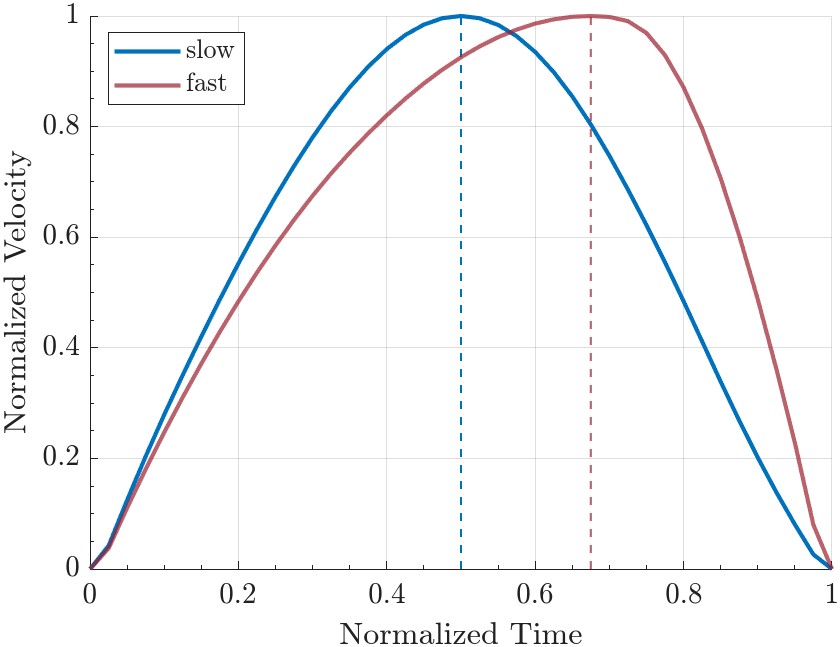}
    \caption{Normalized end effector velocity profiles for slow and fast planned reaching motions. The fast maximum velocity was 1.05 m/s and occurred at a normalized time of 0.68, while the slow maximum velocity was 0.76 m/s and occurred at a normalized time of 0.50.}
    \label{fig:VelocityFeedforward}
\end{figure}

\begin{figure}[h]
    \centering
    \includegraphics[width=1\linewidth]{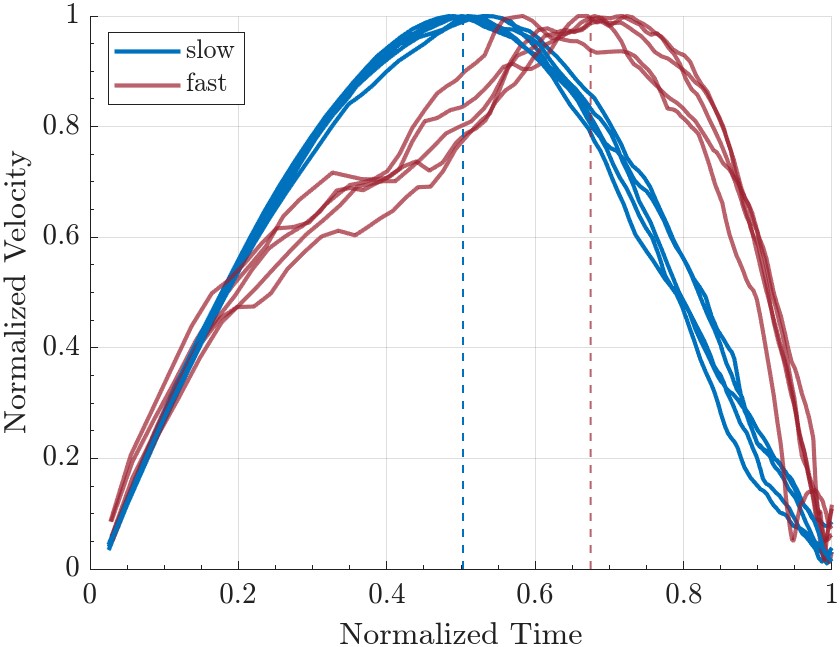}
    \caption{Normalized end effector velocity profiles for slow and fast MPC reaching simulations. The fast maximum velocity was 2.04 m/s and occurred at a normalized time of 0.67, while the slow maximum velocity was 0.76 m/s and occurred at a normalized time of 0.50.}
    \label{fig:VelocityMPC}
\end{figure}

To tune and validate our offline planning model, we first performed a series of reaching simulations with different combinations of $k_u$, $k_t$, $\mathbf{x}_0$, $\mathbf{p}_{\text{target}}$, and $\sigma_{\text{target}}$. Target width $W$ was encoded as $\sigma_{\text{target}}$, effectively defining the target as the 95\% confidence ellipse for final end effector position. For each simulation, we recorded the reaching duration and the value and time of the maximum velocity of the end effector mean trajectory. These values were then compared to experimental data from \cite{fitts_law_exp_data} and ideal parameters were chosen based on similarity to the data. The same procedure was followed for the feedback MPC model, except that the actual end effector trajectory was recorded and each trial was repeated several times and averaged to reduce the effect of noise.
Normalized velocity profiles for a slow and a fast trial of the offline model are shown in Figure \ref{fig:VelocityFeedforward}, and the same is shown for the feedback MPC model in Figure \ref{fig:VelocityMPC}.


In the feed forward case, maximum velocities and normalized times of both slow and fast trajectories were within the experimental ranges reported in \cite{soechting_target_size}.
When comparing maximum velocities between large and small targets, the offline planning model shows a normalized time delay of 0.18 between the two, which is larger but still within the same order of magnitude as the average delay from \cite{soechting_target_size} ($\approx 0.08$). 

In the MPC case, the maximum velocity of the fast trajectory was slightly outside of the experimental range, but all other metrics were within the ranges reported in \cite{soechting_target_size}.
The normalized time delay between large and small targets was 0.17, which is similar to the offline planning model.

Qualitatively, these velocity profiles reproduce several trends found in experimental data.
Existing studies show that normalized velocity profiles will be symmetric for medium velocities, shift to the left for slow movements, and shift to the right for fast movements \cite{asymmetric_vel_acc}\cite{human_vel_curves}. 
Our simulations produce a similar trend, with the normalized velocity profile becoming less symmetric as movement duration decreases.
these results are more realistic than a pure motor noise model, which displays primarily symmetric velocity profiles \cite{signal_dependent_motor_noise}; they are also more realistic than a torque model, which shows poorer correlation with experimental data \cite{original_paper_high_fidelity}.
Although the model fails to replicate a shift of the normalized velocity peak to the left at the lowest speed ranges, it importantly replicates the trend of slower movements having earlier peak velocities, which is a key feature of the experimental data.
From these results we can conclude that although our underlying dynamical model is not as accurate as the high fidelity model used in \cite{original_paper_high_fidelity}, both the offline planning and online MPC reaching models show a level of realism that make them useful for studying the speed-accuracy tradeoff in reaching movements.

\subsection{Fitt's Law Experiment}


In order to determine Fitt's Law parameters for each model, we performed a series of simulations with varying index of difficulty (ID) and recorded the movement duration for each trial.
We varied both the distance to the target and the target width to obtain a range of ID values from 2.14 to 3.22, calculated using eq. \ref{eq:fitts_law}.
The target radius was varied from 0.03 m to 0.04 m, and the target distance was varied from 0.35 m to 0.60 m.
All other parameters were held constant across trials, and the same set of parameters was used for both the offline planning and online MPC models.
$N = 40$ nodes were used, with $t_{\text{iter}} = 0.1$ s, $k_u = 1$, and $k_t = 100$. 

\begin{figure}[h]
    \centering
    \includegraphics[width=0.8\linewidth]{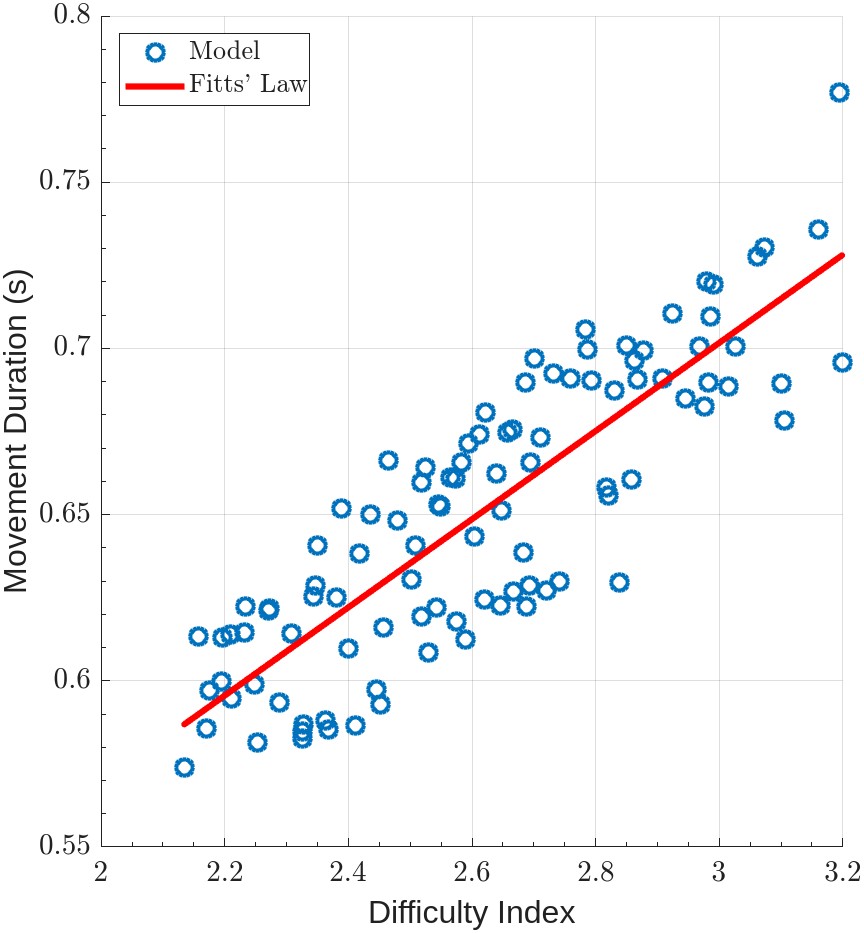}
    \caption{Speed-accuracy tradeoff for the offline planning model. Fitts' Law parameters are $a = 0.25$ and $b = 0.19$, with $R^2 = 0.81$.}
    \label{fig:FittsLawSingle}
\end{figure}

\begin{figure}[h]
    \centering
    \includegraphics[width=0.8\linewidth]{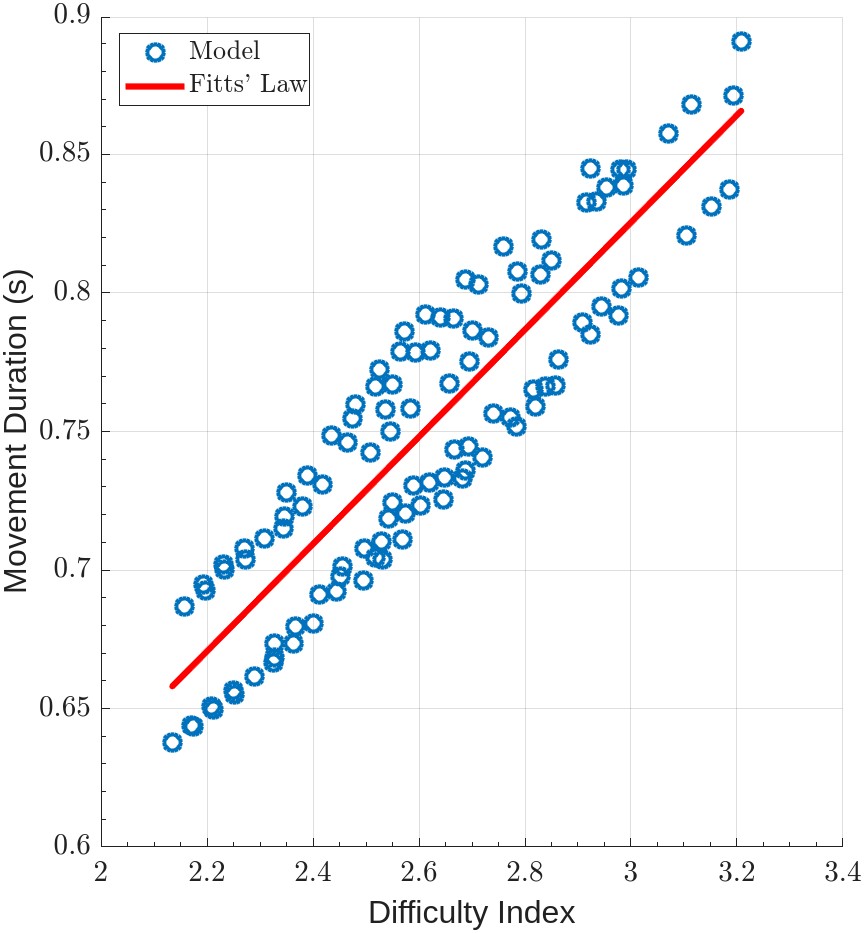}
    \caption{Speed-accuracy tradeoff for the feedback MPC model. Fitts' Law parameters are $a = 0.30$ and $b = 0.13$, with $R^2 = 0.72$.}
    \label{fig:FittsLawMPC}
\end{figure}

Once all trials were complete, a least-squares fit was performed on the data from each model in order to determine the $a$ and $b$ parameters for Fitts' Law. Fit lines for the offline planning and online MPC models are shown in Figures \ref{fig:FittsLawSingle} and \ref{fig:FittsLawMPC}, respectively.   

Both models produce $a,b$ values that fall within the experimentally observed ranges of $a \in [0.0047, 0.5239]$ and $b \in [0.0393,0.1987]$ reported in \cite{fitts_law_exp_data}, and their $R^2$ values indicate good agreement with Fitts' Law.
The offline planning model produces a higher slope and lower offset than the online MPC model, meaning the feedback controller was able to execute movements more quickly than planned, and the actual movement durations were less sensitive to changes in ID than the planned durations. This suggests that as the controller replans throughout the trajectory, it is able to find more optimal trajectories than the original offline plan.
It is also worth noting that the lower $R^2$ value of the MPC data indicates a higher variability with respect to Fitts' law, which is likely due to the stochastic nature of the MPC model.


Since our models are both able to partially replicate the kinematic profiles of real reaching experiments, and the movement duration predicted by our simulations can be predicted by Fitts' Law, our model is reasonably effective at representing the speed-accuracy tradeoff phenomenon.
Our model surpasses the capabilities of a simple torque model, indicating that trajectory optimization while applying signal dependent motor noise to muscle activation is an effective and promising method for improving the modeling of human reaching.
Both noise and trajectory optimization appear to be valid explanations for speed-accuracy tradeoff, and they are not mutually exclusive.

\section{Limitations and Future Work}

One limitation of this work is the lack of a more complex arm model. While the 2D model was sufficient to demonstrate the speed-accuracy tradeoff, a more complex model like the one used in \cite{original_paper_high_fidelity} could provide more biologically realistic data and results, and would allow us to interpet the details of our simulated results rather than simply looking at high level trends.

Another avenue for future work would be to explore and compare different optimization methods for solving the nonlinear optimization problem. While IPOPT produced sufficient results for our purposes, it failed to converge in many cases which restricted the range of reaching situations we could simulate. Additionally, the deterministic nature of IPOPT does not allow for stochastic variation in trajectory solutions given the same set of parameters, which is an important feature of human reaching. To address this, stochastic constrained optimization methods could be explored to make the model more biologically realistic.

Finally, a more realistic model could be created by adding sensor noise and feedback delay to the MPC model. This would more accurately model the full human motor control feedback loop, where the current arm state cannot be sensed perfectly or acted upon instantaneously.

\bibliography{egbib}
\bibliographystyle{IEEEtran}
\end{document}